\documentclass[amsmath,amssymb,twocolumn]{revtex4}
\usepackage{graphicx}
\usepackage{amscd,amsmath,amsthm,amsfonts,amssymb}

\def\sech{{\rm sech}}

\def\ri{{\rm i}}

\def\re{{\rm e}}

\begin{document}
\title{Nonlinear dynamics of wave packets in $\mathcal{PT}$-symmetric optical lattices near the phase transition point}
\author{Sean Nixon,$^{1}$ Yi Zhu,$^2$  and Jianke Yang$^{1}$}

\address{
$^1$Department of Mathematics and Statistics, University of Vermont, Burlington, VT 05401, USA
\\
$^2$Zhou Pei-Yuan Center for Applied Mathematics, Tsinghua
University, Beijing, China }

\begin{abstract}
Nonlinear dynamics of wave packets in $\mathcal{PT}$-symmetric
optical lattices near the phase-transition point are analytically
studied. A nonlinear Klein-Gordon equation is derived for the
envelope of these wave packets. A variety of novel phenomena known
to exist in this envelope equation are shown to also exist in the
full equation including wave blowup, periodic bound states and
solitary wave solutions.
\end{abstract}

\maketitle

Linear Schr\"odinger operators with complex but parity-time
($\mathcal{PT}$)-symmetric potentials have the unintuitive property
that their spectra can be completely real \cite{Bender1998}. This
phenomenon was first studied in quantum mechanics where a real
spectrum is required to guarantee real energy levels. The same
phenomenon was later investigated in paraxial optics, where
$\mathcal{PT}$-symmetric potentials could be realized by employing
symmetric index guiding and an antisymmetric gain/loss profile
\cite{Christodoulides2007,Musslimani2008b}. In this optical setting,
$\mathcal{PT}$ potentials have been experimentally realized
\cite{Guo2009, Segev2010}. In temporal optics,
$\mathcal{PT}$-symmetric lattices have been experimentally obtained
as well \cite{PT_lattice_exp}. So far, a number of novel physical
phenomena in optical $\mathcal{PT}$ systems have been reported,
including phase transition ($\mathcal{PT}$-symmetry breaking),
nonreciprocal Bloch oscillation, unidirectional propagation,
distinct pattern of diffraction, formation of soliton families, and
so on \cite{Guo2009, Segev2010,PT_lattice_exp,
Musslimani2008,Longhi_2009,Musslimani_diffraction_2010,Christodoulides_uni_2011,Nixon2012}.
The search of additional new behaviors in optical $\mathcal{PT}$
systems is still ongoing.

In this Letter, we analytically study nonlinear dynamics of wave
packets in $\mathcal{PT}$-symmetric optical lattices near the
phase-transition point (where bandgaps close and Bloch bands
intersect transversely like the letter `x'). A nonlinear
Klein-Gordon equation is derived for the envelope of these wave
packets near the band intersection. Based on this envelope equation,
we predict a variety of novel phenomena such as wave splitting, wave
blowup, periodic bound states and solitary wave states. We further
show these predicted phenomena occur in the full model as well.

The paraxial model for nonlinear propagation of light beams in
$\mathcal{PT}$-symmetric optical lattices is
\begin{equation}
\ri \Psi_z + \Psi_{xx} + V(x)\Psi + \sigma |\Psi|^2 \Psi = 0,
\label{Eq:NLS}
\end{equation}
where $z$ is the propagation axis, $x$ is the transverse axis,
\begin{equation}
V(x) = V_0^2 \left[ \cos(2x) + \ri W_0 \sin(2x) \right]
\end{equation}
is a $\mathcal{PT}$-symmetric potential, $V_0^2$ is the potential
depth, $W_0$ is the relative gain/loss strength, and $\sigma=\pm 1$
is the sign of nonlinearity. All variables are non-dimensionalized.

\begin{figure}[htb]
\includegraphics[width=8.3cm]{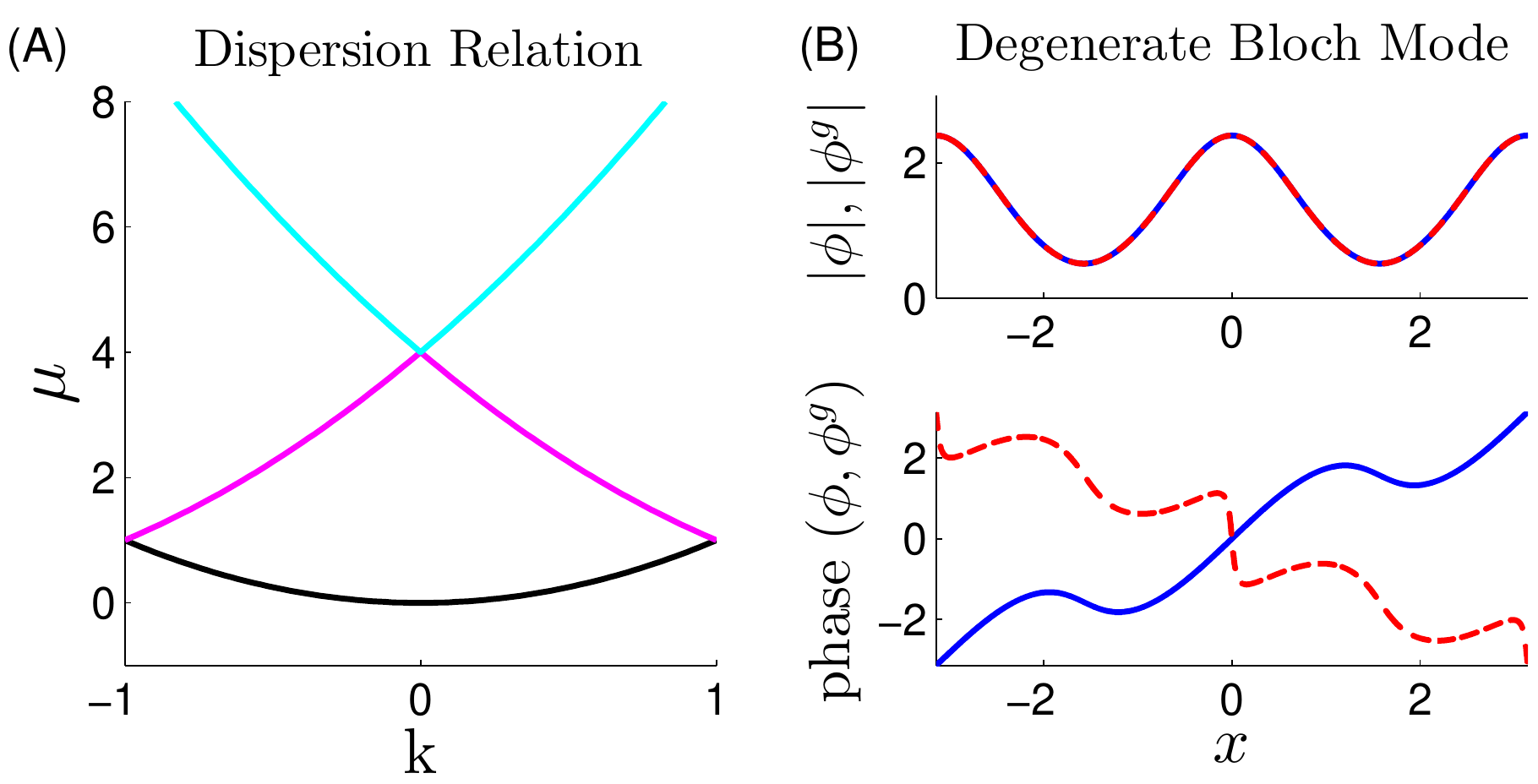}

\vspace{0.1cm}
\includegraphics[width=8.3cm]{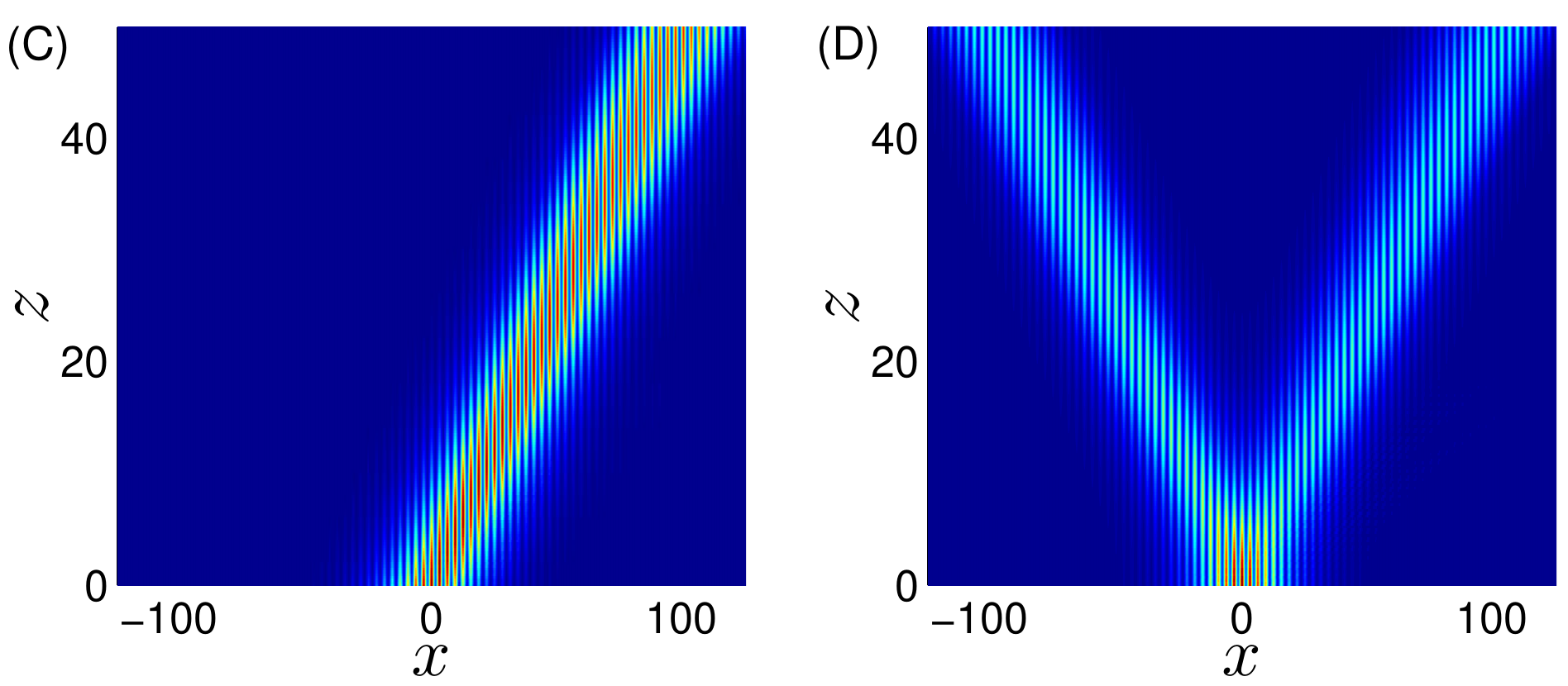}
\caption{(A) Diffraction relation; (B) Magnitude
and phase of eigenfunction $\phi$ (solid blue)
and generalized eigenfunction $\phi^g$ (dashed red) for $\mu=1$ and $V_0=\sqrt{6}$;
(C) Linear unidirectional wavepacket and
(D) linear wavepacket splitting in Eq. (\ref{Eq:NLS}) at the phase-transition point $W_0=1$.
} \label{Fig:Diffraction}
\end{figure}

First we discuss the linear diffraction relation of Eq.
(\ref{Eq:NLS}) at the phase-transition point $W_0=1$
\cite{Musslimani2008,Nixon2012}. In this case, the linear equation
of (\ref{Eq:NLS}) can be solved exactly \cite{Nixon2012}. Its Bloch
solutions are
\begin{equation}  \label{e:Bloch}
\Psi_\pm (x, z; \mu ) = I_{\pm (k+2m)}\left( V_0 \re^{\ri x} \right) \re^{-\ri \mu z},
\end{equation}
where $I_k$ is the modified Bessel function, $\mu = (k + 2m)^2$ is
the diffraction relation, $k$ is in the first Brillouin zone $k\in
[-1,1]$, and $m$ is any nonnegative integer. This diffraction
relation is depicted in Fig. \ref{Fig:Diffraction}(A), where
different colors indicate different Bloch bands. At $k = 0$ and $\pm
1$, adjacent Bloch bands intersect each other transversely like the
letter `x'. Due to this intersection (degeneracy), wave packets near
these points will exhibit novel dynamics which we will reveal next.

At these intersection points, $\Psi_+ = \Psi_-$, thus Bloch
solutions (\ref{e:Bloch}) are degenerate and  $2\pi$-periodic in
$x$. Posed as an eigenvalue problem for $\Psi = \phi(x) \re^{-\ri
\mu z}$ in the linear equation of (\ref{Eq:NLS}), we get $L \phi  =
-\mu \phi$, where $L \equiv \partial_{xx} + V_0(x)$, and $V_0(x)$ is
the $\mathcal{PT}$ lattice at the phase-transition point $W_0=1$.
Then at these band-intersection points, the eigenvalues are $\mu =
n^2$, where $n$ is any positive integer. The corresponding
eigenfunctions are
\begin{equation}  \label{e:phi}
\phi(x)=I_n \left( V_0 \re^{\ri x} \right)
=\sum_{j=0}^\infty \frac{(V_0e^{ix}/2)^{2j+n}}
{j!(j+n)!}.
\end{equation}
These eigenvalues  $\mu = n^2$ all have geometric multiplicity 1 and
algebraic multiplicity 2, thus there exist a generalized
eigenfunction $\phi^g$ satisfying
\begin{equation}  \label{d:phig}
\left( L+\mu \right) \phi^g = \phi.
\end{equation}
For $V_0=\sqrt{6}$ and $n=1$, this eigenfunction and generalized
eigenfunction are plotted in Fig.~\ref{Fig:Diffraction}(B).

\small
\normalsize

Now we consider nonlinear dynamics of wave packets near these
band-intersection points. For convenience, we first take the
$\mathcal{PT}$ lattice to be exactly at the phase-transition point
(i.e., $W_0=1$). Generalization to lattices near the
phase-transition point will be made afterwards.

Nonlinear wave-packet solutions near a band-intersection point can
be expanded into a perturbation series
\begin{equation}
\Psi =  \re^{-\ri \mu z} [\epsilon A(X,Z) \phi(x) + \epsilon^2 \psi_1 + \epsilon^3 \psi_2+ \ldots],
\label{Eq:Expansion}
\end{equation}
where $\mu=n^2$ is the propagation constant at the band
intersection, $\phi(x)$ is the degenerate Bloch mode given in
Eq.~(\ref{e:phi}), $A(X,Z)$ is the envelope of this Bloch mode,
$X=\epsilon x$, $Z=\epsilon z$ are slow variables, and
$0<\epsilon\ll 1$ is a small positive parameter. Substituting the
above perturbation series into the original equation (\ref{Eq:NLS}),
this equation at $O(\epsilon)$ is automatically satisfied. At order
$\epsilon^2$ we have
\begin{equation}
(L+\mu)\psi_1= -\ri A_Z \phi - 2 A_X \phi_x.
\end{equation}
The solution to this equation is
\begin{equation}  \label{e:psi1}
\psi_1 = -\ri A_Z \phi^g - 2 A_X \phi^d,
\end{equation}
where $\phi^g$ is the generalized Bloch mode defined in
(\ref{d:phig}), and
\begin{equation}
\phi^d = (L+\mu)^{-1} \phi_x.     \label{Eq:Phi1}
\end{equation}
This $\phi^d$ solution exists and can be determined by Fourier
series \cite{Nixon2012}.

At O($\epsilon^3$) we get
\begin{eqnarray}
\hspace{-1cm} (L+\mu) \psi_2  =
-A_{ZZ} \phi^g + A_{XX} \left(4\phi^d_x - \phi\right)    \nonumber
\\  +\ri 2A_{ZX} \left( \phi^d +\phi^g_x\right)- \sigma |A|^2A |\phi|^2 \phi.  \hspace{-1cm}
\end{eqnarray}
The solvability condition of this equation is that its right hand
side be orthogonal to the adjoint homogeneous solution $\phi^*$.
Using Fourier expansions of solutions $\phi^g$ and $\phi^d$ (see
\cite{Nixon2012}), this solvability condition then yields the
following nonlinear Klein-Gordon equation for the envelope function
$A(X,Z)$:
\begin{equation}
A_{ZZ} - 4n^2 A_{XX} + \gamma  |A|^2A = 0,
\label{Eq:NLW}
\end{equation}
where
\begin{equation}  \label{e:gamma}
\gamma = (-1)^{n+1}\frac{2\sigma n^2}{\pi} \int_{-\pi}^\pi |\phi|^2 \phi^{2} dx.
\end{equation}
In view of the formula (\ref{e:phi}) for $\phi$, it is easy to see
that $\mbox{sgn}(\gamma)=(-1)^{n+1}\sigma$. For the values of
$V_0=\sqrt{6}$ and $n=1$ which we will use in later numerical
simulations, $\gamma\approx  11.0430\sigma$.

Now we extend the above envelope equation to the case where the
$\mathcal{PT}$ lattice is near the phase-transition point (i.e.,
$W_0\sim 1$). Following similar perturbation analysis, we find that
when $\mu=n=1$ (the lowest band-intersection point) and
$W_0=1-c\hspace{0.04cm} \epsilon^2$, the envelope is governed by a
slightly more general nonlinear Klein-Gordon equation,
\begin{equation}
A_{ZZ} - 4n^2A_{XX}  +\alpha A+ \gamma  |A|^2A = 0,
\label{Eq:NLWn}
\end{equation}
where $\gamma$ is as given in (\ref{e:gamma}), $\alpha=c V_0^4/2$,
and the $\psi_1$ solution in (\ref{Eq:Expansion}) is still given by
Eq. (\ref{e:psi1}). When $\mu=n^2$ with $n>1$ and
$W_0=1-c\hspace{0.04cm} \epsilon$, the envelope equation
(\ref{Eq:NLWn}) will contain an additional term proportional to
$iA_Z$. But this $iA_Z$ term can be eliminated through a gauge
transformation $A\to A \hspace{0.04cm} \re^{-\ri c V_0^4/4(n^2-1)}$,
and the transformed equation remains the same as (\ref{Eq:NLWn}),
except that $\alpha=c^2V_0^8/64$ for $n=2$ and $\alpha=0$ for $n>2$.
If $c=0$ (i.e., at the phase-transition point), then $\alpha=0$,
hence Eq. (\ref{Eq:NLWn}) reproduces Eq. (\ref{Eq:NLW}) as a special
case. When $n>1$, the $\psi_1$ solution (\ref{e:psi1}) will also
contain an additional term proportional to $icA$.

The nonlinear Klein-Gordon equation (\ref{Eq:NLWn}) is second-order
in $Z$, which means that two initial conditions, $A(X, 0)$ and
$A_Z(X, 0)$, are needed. These two initial conditions can be
obtained from the initial conditions of the perturbation series
(\ref{Eq:Expansion}) at orders $\epsilon$ and $\epsilon^2$, i.e.,
from the initial envelope $A(X,0)$ as well as $\psi_1|_{z=0}$. This
$\psi_1|_{z=0}$ generally contains many eigenmodes of the operator
$L$, but only the $\phi^g$ component in it affects the dynamics of
envelope $A$. If we denote $B(X)$ as the envelope function of the
$\phi^g$ component in $\psi_1|_{z=0}$, then by projecting
$\psi_1|_{z=0}$ (such as from (\ref{e:psi1})) onto $\phi^g$, we find
that
\begin{equation}
B(X) = -\ri  \left[ A_Z(X,0) + 2 n A_X(X,0) \right]. \label{Eq:BEnvelope}
\end{equation}
This relation holds for all $n$ values. Thus, initial conditions for
the original $\mathcal{PT}$ model (\ref{Eq:NLS}) and those for the
envelope equation (\ref{Eq:NLWn}) can be related as
\begin{equation}
\Psi(x,0)=\epsilon A(X,0)\phi(x) + \epsilon^2 B(X) \phi^g(x).
\end{equation}
This initial-value connection will be used in our numerical
simulations later.

Now we examine the envelope dynamics in the nonlinear Klein-Gordon
equation (\ref{Eq:NLWn}), and show that the corresponding wavepacket
dynamics occurs in the original $\mathcal{PT}$ model (\ref{Eq:NLS})
too. In this discussion, we take $n=1$, $V_0=\sqrt{6}$ and
$\epsilon=0.1$ for definiteness.

First we consider the linear Klein-Gordon equation (\ref{Eq:NLWn})
at the phase-transition point, i.e., $A_{ZZ} - 4n^2A_{XX}=0$. This
is the familiar second-order wave equation. It admits unidirectional
wave solutions $F(X-2nZ)$ as well as bidirectional wave solutions
$F(X-2nZ)+G(X+2nZ)$. In the original linear $\mathcal{PT}$ model
(\ref{Eq:NLS}), we find that the corresponding wavepacket solutions
also exist. Examples are displayed in
Fig.~\ref{Fig:Diffraction}(C,D). Note that the wave splitting in
Fig.~\ref{Fig:Diffraction}(D) is manifestation of the `x'-shaped
diffraction structure at the band-intersection point, while the
unidirectional propagation in Fig.~\ref{Fig:Diffraction}(C) occurs
for $B(X) = 0$.

\small
\begin{figure}[htb]
\includegraphics[width=8.1cm]{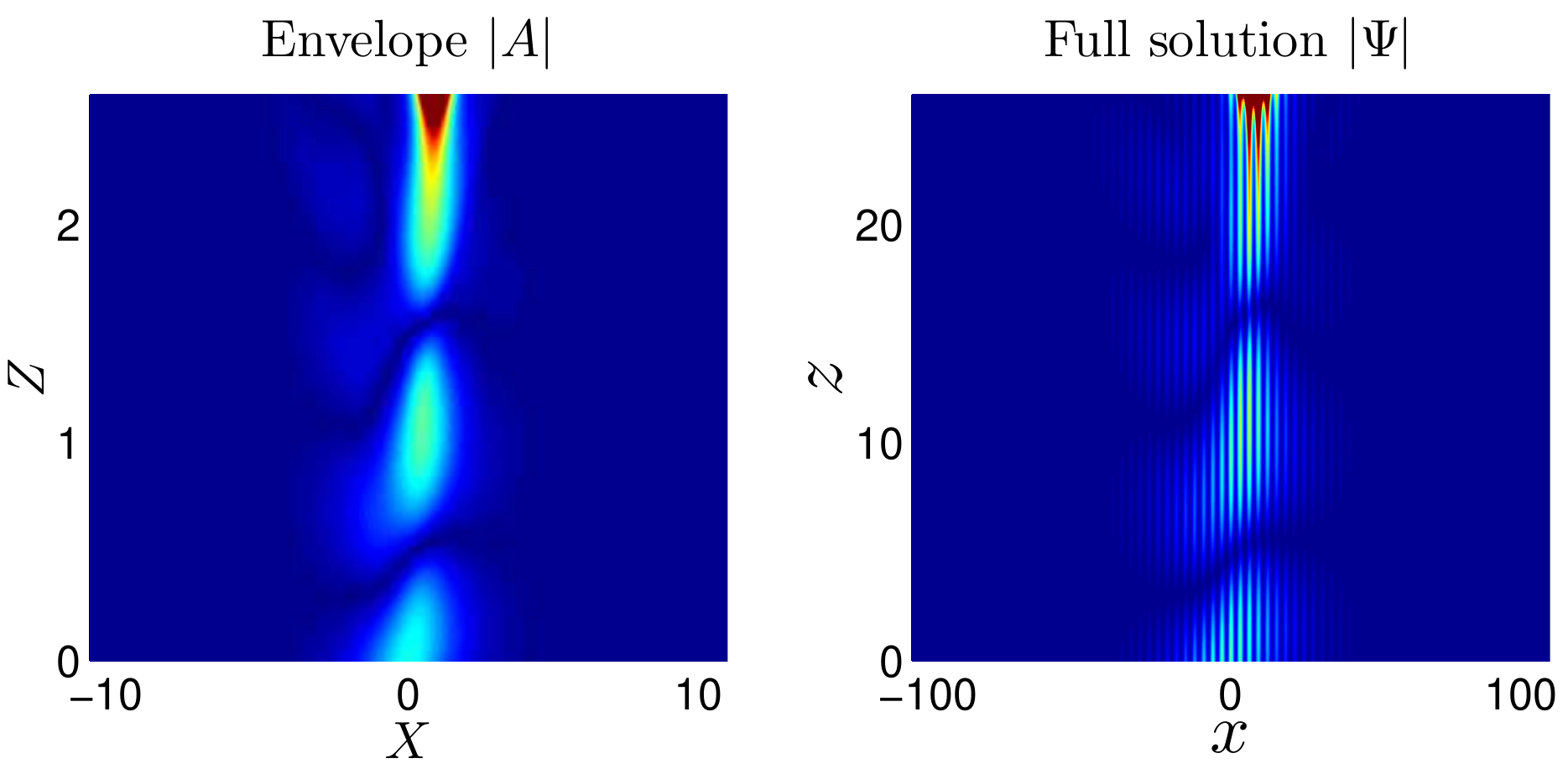}

\vspace{0.1cm}
\includegraphics[width=8.1cm]{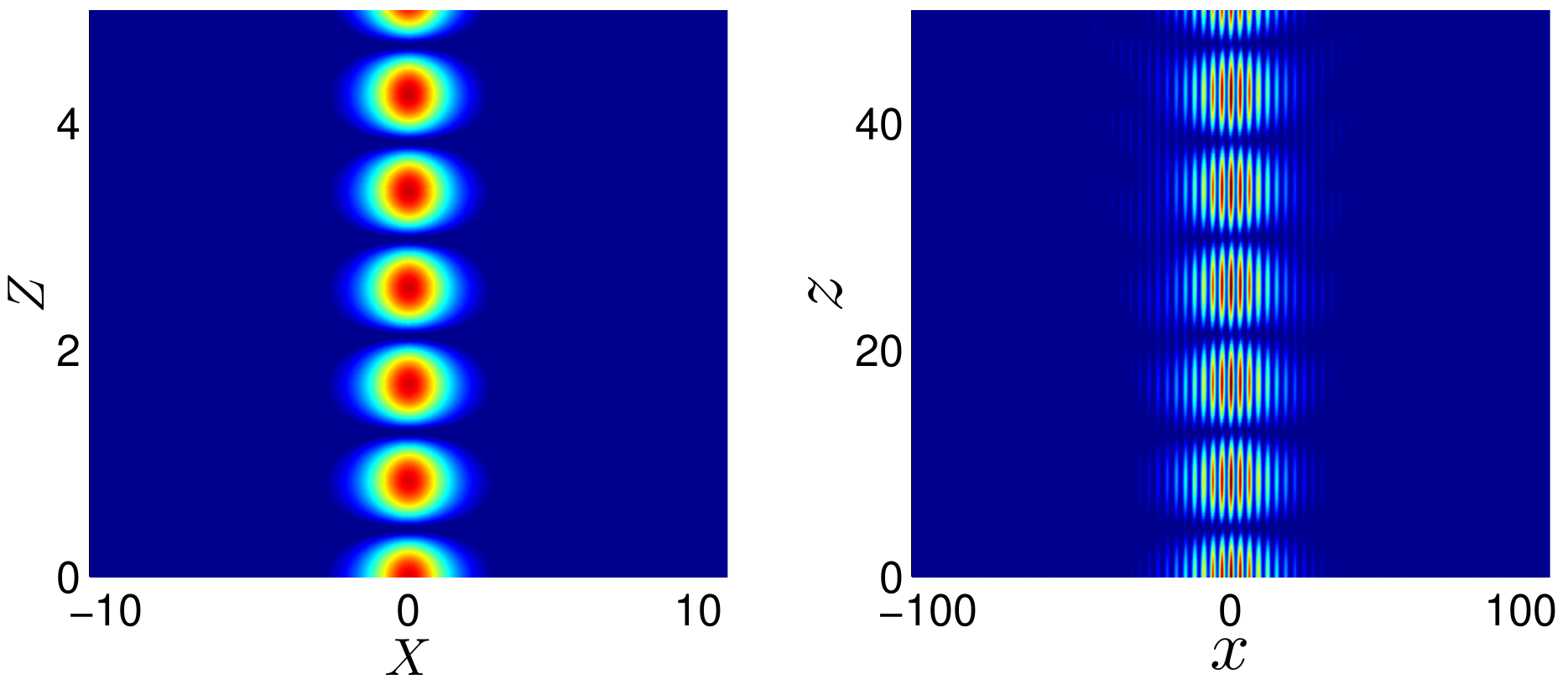}
\caption{Nonlinear wavepacket solutions below the phase-transition
point under self-defocusing nonlinearity.
Top: a blowup solution; bottom: a periodic bound state. Left:
envelope solutions $|A|$; right: full solutions $|\Psi|$.
} \label{Fig:NLExample}
\end{figure}
\normalsize

Next we consider envelope solutions in the nonlinear Klein-Gordon
equation (\ref{Eq:NLWn}) below the phase-transition point, i.e.,
$W_0<1$, or $\alpha>0$ (above the phase-transition point,
infinitesimal linear waves are unstable, thus it is not pursued).
When the nonlinearity is self-defocusing ($\sigma=-1$),
$\gamma\approx  -11.0430<0$. In this case, envelope solutions can
blow up to infinity in finite distance \cite{JohnBook}. One such
example is shown in Fig.~\ref{Fig:NLExample} (upper left panel) for
$c=1$ ($\alpha=18$) and initial conditions
$A(X,0)=1.2\hspace{0.04cm} \sech(X)$, $A_Z(X,0)=-2 A_X(X,0)$ ($B(X)
= 0$). In the full nonlinear $\mathcal{PT}$ model (\ref{Eq:NLS}), we
have found similar blowup solutions which are displayed in Fig.
\ref{Fig:NLExample} (upper right panel). This solution-blowup under
self-defocusing nonlinearity is very surprising.
Note that in the full model (\ref{Eq:NLS}), our asymptotic envelope
approximation breaks down as the singular (blowup) point is
approached. In this case, the amplitude of the full-model solution
eventually saturates, but its power still grows unbounded
\cite{Nixon2012}.

Under self-defocusing nonlinearity, the envelope equation
(\ref{Eq:NLWn}) also admits $X$-localized and $Z$-periodic bound
states. One example with $c=1$ (below the phase-transition point) is
shown in Fig.~\ref{Fig:NLExample} (lower left panel). The initial
condition for this solution is $A(X,0)=\sech(X)$ and $A_Z(X,0)=0$
$(B(X) =- 2i A_X(X,0))$.  In the full nonlinear $\mathcal{PT}$ model
(\ref{Eq:NLS}), we have found similar periodic bound states as well,
see Fig. \ref{Fig:NLExample} (lower right panel).

Under self-defocusing nonlinearity and below the phase-transition
point, the envelope equation (\ref{Eq:NLWn}) also admits stationary
solitary waves $A(X,Z) =F(X)\re^{\ri \omega Z}$ when $\omega$ lies
inside the bandgap $-\sqrt{\alpha}<\omega< \sqrt{\alpha}$. These
solitary envelope solutions correspond to the solitons of the full
nonlinear $\mathcal{PT}$ model (\ref{Eq:NLS}) reported in
\cite{Musslimani2008,Nixon2012}.

If the nonlinearity is self-focusing ($\sigma=1$), envelope
solutions do not blow up, periodic bound states cannot be found, and
stationary solitary waves do not exist in the envelope equation. In
this case, nonlinear diffracting solutions similar to the linear
diffracting pattern reported in \cite{Musslimani_diffraction_2010}
exist. In addition, solutions periodic in both $X$ and $Z$ can be
found.
%

In summary, we have shown that the nonlinear Klein-Gordon equation
governs the envelope dynamics of wavepackets in $\mathcal{PT}$
lattices near the phase-transition point. We have also shown that a
variety of novel phenomena in this envelope equation (such as wave
blowup and periodic bound states) occur in the full $\mathcal{PT}$
model too. These findings open new possibilities for wave
engineering in $\mathcal{PT}$ lattices.

This work is supported in part by AFOSR.

\end{document}